\documentstyle[aasms4,12pt]{article}
\begin{document}
\title{Periodic X-ray Emission from the O7~V Star $\theta^1$~Orionis~C}
 
\author{Marc Gagn\'e\altaffilmark{1}, Jean-Pierre Caillault\altaffilmark{2}, 
John R. Stauffer\altaffilmark{3}, and Jeffrey L. Linsky\altaffilmark{1}}

\altaffiltext{1}{JILA, University of Colorado, Boulder, CO 80309-0440}
\altaffiltext{2}{Department of Physics and Astronomy, University of Georgia, Athens, GA 30602}
\altaffiltext{3}{Harvard-Smithsonian Center for Astrophysics, 60 Garden Street, Cambridge, MA 02138}
\received{1996 November 26}
\accepted{1997 January 13}
\slugcomment{to appear in The Astrophysical Journal Letters}
\righthead{GAGN\'E. ET AL.}
\lefthead{PERIODIC X-RAYS FROM $\theta^1$~ORI~C}

\begin{abstract}

We report the discovery of large-amplitude, periodic X-ray emission
from the O7 V star $\theta^1$~Orionis~C, the central star of the Orion
Nebula.  Ten {\it ROSAT} HRI snapshots of the Trapezium cluster taken
over the course of 21 days show that the count rate of
$\theta^1$~Ori~C varies from 0.26 to 0.41 counts~s$^{-1}$
with a clear 15-day period.  The soft X-ray variations have the same
phase and period as H$\alpha$ and \ion{He}{2} $\lambda 4686$ variations
reported by Stahl et al., and are in anti-phase with the
\ion{C}{4} and \ion{Si}{4} ultraviolet absorption features.
We consider five mechanisms which might explain
the amplitude, phase, and periodicity of the X-ray variations:
(1)~colliding-wind emission with an unseen binary companion,
(2)~coronal emission from an unseen late-type pre-main--sequence star,
(3)~periodic density fluctuations,
(4)~absorption of magnetospheric X-rays in a corotating wind, and
(5)~magnetosphere eclipses.
The {\it ROSAT} data rule out the first three scenarios,
but cannot rule out either of the latter two which
require the presence of an extended magnetosphere,
consistent with the suggestion of Stahl et al.
that $\theta^1$~Ori~C is an oblique magnetic rotator.
As such, $\theta^1$~Ori~C may be the best example of a
high-mass analog to the chemically peculiar, magnetic Bp stars.
 
\end{abstract}
 

\section{Introduction}

$\theta^1$~Orionis~C (HD~37022=HR~1895) is the central star of the Trapezium cluster
and the principal source of ultraviolet photons illuminating the Orion Nebula
(M~42). $\theta^1$~Ori~C also illuminates the population of proplyds 
in its vicinity (O'Dell \& Wen 1994), providing some of the most direct evidence 
yet for circumstellar disks around low-mass pre-main--sequence (PMS) stars.
It is classified as O7~V (Conti \& Leep 1974) but is 
spectroscopically variable (Conti 1972; Walborn 1981).
Conti \& Alschuler (1971) and Conti (1972) also found variable inverse P Cygni
emission profiles in the \ion{He}{2}~$\lambda 4686$ line.
Recently, Stahl et al. (1993) discovered
that the long-known H$\alpha$ and \ion{He}{2}~$\lambda 4686$ emission variations
on $\theta^1$~Ori~C exhibit a strict, stable 15.4-day periodicity, presumably
the rotation period of the star.
They suggested that $\theta^1$~Ori~C may be an oblique magnetic
rotator akin to magnetic chemically peculiar B stars like the 
He-strong star $\sigma$~Ori~E (B2~Vpe) where the emission
lines are likely to be formed in the magnetosphere above 
the magnetic equator (Shore \& Brown 1990).

Walborn \& Nichols (1994) and Stahl. et al. (1996) subsequently 
found \ion{C}{4} $\lambda\lambda 1548, 1551$ absorption
variations at high velocities that are consistent with the 
H$\alpha$ period and concluded that
$\theta^1$~Ori~C possesses a corotating wind.
The \ion{C}{4} absorption strength is minimum
when the H$\alpha$ and \ion{He}{2}~$\lambda 4686$ emission
is maximum, i.e., when the wind in our line of sight 
is weakest.  From the phase difference between the optical 
emission and the UV absorption-line strength,
Stahl et al. (1996) infer a rotational inclination
$i \sim 45^{\circ}$ and a magnetic obliquity $\beta \sim 45^{\circ}$.

X-ray variability of $\theta^1$~Ori~C was first reported by 
Ku, Righini-Cohen, \& Simon (1982) based on {\it Einstein} HRI 
observations of the Trapezium.  Three deep {\it ROSAT} HRI
exposures obtained in 1991 October, 1992 March, and 1992 September
show that $\theta^1$~Ori~C
varied from 0.29 to 0.45 counts~s$^{-1}$.
When folded with the ephemeris of Stahl et al. (1993), the high and low X-ray
count rates corresponded to phases of maximum and minimum H$\alpha$ emission,
respectively (Caillault, Gagn\'e, \& Stauffer 1994).

\section{Observations and Results}

During the period 1995 September 1--21, the {\it ROSAT} HRI 
(cf. David et al. 1996) obtained
10 observations of the Trapezium each separated by approximately two days
and each consisting of two satellite orbits.
A light curve for $\theta^1$~Ori~C was generated by summing
counts within a $10\arcsec$ (90\% power) radius circle centered on the
emission peak, subtracting a local background, and dividing the net counts
by the dead-time corrected exposure time for each orbit.
The source region dimensions were chosen to exclude essentially
all counts from other Trapezium sources.  
In Figure~1, we plot the HRI light curve of $\theta^1$~Ori~C.
Images of individual orbits revealed that for 7 of the 20 orbits, the standard 
processing software was unable to generate an adequate aspect solution.
Consequently, point sources in these 7 images appear to be smeared
along the SE--NW axis. For these orbits, source counts 
were extracted from an ellipse with major and minor
axes of $13.1\arcsec$ and $9.2\arcsec$, respectively, and position angles
ranging from $40^{\circ}$--$50^{\circ}$.
These points are indicated in Fig.~1 with crosses.

\placefigure{figure_1}
  
In Fig.~1, the upper axis indicates the corresponding phase
using the ephemeris of Stahl et al. (1996),
where $P = 15.422$~d, $M\!J\!D = 48832.5$, and phase 0.0
correspond to maximum H$\alpha$ emission.  
Also plotted are the 1991 Oct 2, 1992 Mar 22, and 1992 Sep 14 count rates
and phases (open circles).
Fitting a simple sine curve to the X-ray count rates with 
a period of 15.422~d yields
a low-state count rate of 0.26~counts~s$^{-1}$ and an amplitude
of 0.15~counts~s$^{-1}$.  The fit is plotted in Fig.~1.  
Leaving the period as a free parameter, we find
a period of $16.0 \pm 3.8$~days ($1 \sigma$ error). 
The soft X-ray emission appears to vary in phase with the H$\alpha$
emission and with a very similar period. 

Based on the optical extinction estimates of Breger, Gehrz, \& Hackwell (1981),
the interstellar column density in the $\theta^1$~Ori~C
line of sight is likely to be $(3 \pm 1) \times 10^{21}$~cm$^{-2}$.
Assuming a distance of 440~pc, the
0.1--2.0~keV X-ray luminosity of $\theta^1$~Ori~C at phase $\sim 0.5$ is
$L_{\rm X} \sim 4.6 \times 10^{32}$~ergs~s$^{-1}$
(corrected for interstellar absorption).
If we assume no major changes in the spectral shape at X-ray maximum,
the 0.15~counts~s$^{-1}$ HRI variability amplitude corresponds
to $\Delta L_{\rm X} \sim 2.6 \times 10^{32}$~ergs~s$^{-1}$.

\section{Discussion}

We discuss the X-ray variability in the context of five different models: 
(1)~colliding-wind emission with an unseen lower-mass companion,
(2)~coronal emission from an unseen lower-mass companion,
(3)~periodic density variations, 
(4)~absorption of magnetospheric X-rays in a corotating wind, and
(5)~magnetosphere eclipses.

\subsection{Colliding Winds}

For the first two scenarios, we assume that $\theta^1$~Ori~C has
an unseen PMS companion.  While there is no compelling evidence
that $\theta^1$~Ori~C is a binary, many high-mass stars 
in the Trapezium are spectroscopic binaries (Abt, Wang, \& Cardona 1991).
Stahl et al. (1996) did find that a number of 
$\theta^1$~Ori~C's photospheric lines
show irregular $\sim 2$~km~s$^{-1}$ radial-velocity variations,
suggesting an upper limit to the companion's mass 
$M \sin i \sim 0.27 M_{\odot}$,
assuming an O star mass $M \sim 36 M_{\odot}$ (Howarth \& Prinja 1989).
If 15.4~d is the orbital period, then a lower limit
for the binary separation is $a \sim 85 R_{\odot}$.
In colliding-wind models, the X-rays arise in a shock region where the wind 
of a hot, massive star collides with either the wind or the outer atmosphere 
of the companion.  We calculate the expected X-ray emission from $\theta^1$~Ori~C
via this mechanism by comparing it to the well-studied example of $\gamma$~Velorum,
a Wolf-Rayet binary (O9~I + WC8) in which the WR secondary possesses a very massive 
wind with $\dot{M} \sim 8.8 \times 10^{-5} M_{\odot}$~yr$^{-1}$
and $v_{\rm \infty} \sim 1520$~km~s$^{-1}$. 
Willis, Schild, \& Stevens (1995) predict 
$L_{\rm X} \sim 10^{32}$~ergs~s$^{-1}$ 
resulting from the $\gamma$~Vel colliding-wind shock region.

In the case of $\theta^1$~Ori~C, the wind is variable
and published values of the terminal wind speed are discordant.
Prinja, Barlow, \& Howarth (1990) used the narrow absorption component in the
\ion{C}{4} profile and estimated  $v_{\infty} \sim 510$~km~s$^{-1}$.
Walborn \& Nichols (1994) identified a broad, variable, high-velocity component
in the \ion{C}{4} profile and determined $v_{\rm \infty} \sim 3600$~km~s$^{-1}$. 
In order to maximize the effect of colliding winds, we assume
$\dot{M} \sim 4 \times 10^{-7} M_{\odot}$~yr$^{-1}$
and $v_{\rm \infty} \sim 510$~km~s$^{-1}$ for $\theta^1$~Ori~C; 
in the most optimistic case, the T Tauri companion might have 
$\dot{M} \sim 4 \times 10^{-7} M_{\odot}$~yr$^{-1}$ (Levreault 1988) 
and $v_{\rm \infty} \sim 200$~km~s$^{-1}$ (Natta, Giovanardi, \& Palla 1988).   
Using these wind parameters in eq.~(10) of 
Stevens, Blondin, \& Pollock (1992), we find that the maximum expected 
X-ray luminosity of $\theta^1$~Ori~C from colliding winds would be comparable to 
that observed on $\gamma$~Vel.

Although the wind parameters have been chosen to maximize colliding-wind 
emission, the predicted $L_{\rm X}$ falls short of the observed X-ray variability 
amplitude by a factor of three. Moreover, the collision of such slow winds 
may not produce sufficient shocks to heat the interaction region to $T > 1$~MK.
If instead we use $v_{\rm \infty} \sim 3600$~km~s$^{-1}$ for the terminal wind speed
of $\theta^1$~Ori~C, the predicted X-ray luminosity is a factor of at least 25
lower because of the lower wind density in the interaction region.
In this case, the sharp peak at phase 0.0 predicted by
the model of Willis et al. (1995) is inconsistent with the smoothly varying 
sinusoidal X-ray emission observed on $\theta^1$~Ori~C.
Also, colliding winds cannot account for the smooth variation with phase 
of the equivalent width of the \ion{C}{4} absorption line observed by
Walborn \& Nichols (1994) and Stahl et al. (1996).
It should be noted that the Stevens et al. (1992)
model assumes spherically symmetric mass loss while mass loss from T Tauri stars
is often collimated in bipolar jets (e.g., Shu et al. 1994).  It is also unclear how 
long a massive accretion disk (which drives mass loss in T Tauri stars)
might survive in close proximity to an O7~V star.
While there are many complicating factors one could include in
the colliding-wind scenario, we think it unlikely that these
complications would significantly improve the agreement between
the model and the observations.

\subsection{A Low-Mass Coronal Companion}

Next, we consider coronal X-ray emission from an unseen, PMS
companion whose coronal X-ray emission 
is being eclipsed by the O-star or severely attenuated
by the O-star wind at the 15.422~d orbital period.
Since the amplitude of the HRI variations corresponds to
$\Delta L_{\rm X} \sim 2.6 \times 10^{32}$~ergs~s$^{-1}$,
this must be a lower limit to the companion's X-ray luminosity.
Since the most X-ray luminous
low-mass star in the entire Orion region, P1817,
has $L_{\rm X} \sim 5 \times 10^{31}$~ergs~s$^{-1}$
(Gagn\'e, Caillault, \& Stauffer 1995), it is unlikely that a low-mass star
can account for the observed variations.  Moreover, as has been pointed
out by Stahl et al. (1996), a low-mass companion cannot account for the
luminosity of the H$\alpha$ and \ion{He}{2} $\lambda 4686$ variations.  
Given the the small-amplitude radial-velocity variations, a more massive 
companion (e.g., an O or early-B star) could exist if
the binary orbit were in the plane of the sky ($i \approx 0$);
but, in this case, we do not expect significant X-ray variability.

\subsection{Periodic Density Fluctuations}

To our knowledge, $\zeta$~Puppis (O4~I(n)f) is the only
other candidate O-type magnetic rotator.
Moffat \& Michaud (1981) report periodic H$\alpha$ variations
and propose a rotational period of 5.075~d for $\zeta$~Pup. 
Bergh\"ofer et al. (1996) report 6\% variations in the 0.9--2.0~keV
{\it ROSAT} PSPC flux and a $2\sigma$
peak in the X-ray power spectrum near 16.7~hr.
Contemporaneous H$\alpha$ spectra show profile variations
consistent with the Moffat \& Michaud (1981) period of 5.075~d.
The residual profile variations also indicate a weak peak in the 
H$\alpha$ emission power spectrum near the X-ray period of 16.7~hr.
Bergh\"ofer et al. interpret the X-ray--H$\alpha$ correlation as evidence
for periodic density fluctuations at the base of the wind.
Interestingly, the PSPC time series does not indicate any variability
at the 5.1-d rotation period.
Nonetheless, the periodic H$\alpha$ and ultraviolet variations
and the possible presence of correlated X-ray--H$\alpha$ emission
in $\zeta$~Pup and $\theta^1$~Ori~C is noteworthy.
Bergh\"ofer et al. (1996) interpret the 16.7~hr period 
as pulsations or cyclically repeating azimuthal structures.
They suggest that either of these will produce periodic density variations
at the photosphere which propagate through the wind, producing  
enhanced H$\alpha$, 
and X-ray emission seen from different characteristic heights in the wind.
Bergh\"ofer et al. (1996) propose that the emissions have the same apparent
phase because the phase shift between the X-ray and H$\alpha$ emission
regions is, fortuitously, $\sim 1$.
In the case of $\theta^1$~Ori~C, the sound crossing time through the wind is short 
($t \lesssim 1$~d out to $r \sim 10 R_{\star}$) compared to the 15.4-d period
of $\theta^1$~Ori~C.
Consequently, a density wave resulting from non-radial pulsations
might lead to apparently coherent variations.
However, the lowest non-radial pulsation modes for main-sequence O stars have
periods in the range 0.5--2.0~d (Baade 1986), much shorter than the 
observed 15.4~d period.

\subsection{Absorption in a Corotating Wind}

For the next two scenarios, we assume that the
15.422~d periodicity is the rotational period of a single O star.
The periodic \ion{C}{4} and \ion{Si}{4} profile variations seen
by Walborn \& Nichols (1994) and Stahl et al. (1996)
appear to require a non-spherically symmetric, corotating wind.
The tremendous torque required to maintain a corotating wind
also suggests an extended magnetic field geometry.
As has been pointed out by Stahl et al. and Walborn \& Nichols,
the \ion{C}{4}, H$\alpha$, and \ion{He}{2} variations on $\theta^1$~Ori~C
are reminiscent of the oblique magnetic rotator $\sigma$~Ori~E (B2~Vpe).
On $\sigma$~Ori~E, mass loss occurs preferentially along open magnetic field
lines over the magnetic poles, while, at lower magnetic latitudes, field lines 
close inside the Alfv\'en radius, funneling wind material towards the magnetic
equator. Bolton (1994) suggests that the region of closed magnetic 
field lines (the magnetosphere) extends out to $R \sim 5 R_{\star}$.
If phases of maximum mass loss on $\theta^1$~Ori~C correspond 
to phases of maximum \ion{C}{4} absorption, then we expect one
magnetic pole to pass our line of sight around phase 0.5.
(Stahl et al. 1996).

We suggest that most of $\theta^1$~Ori~C's X-ray emission is
produced in the magnetosphere of an oblique magnetic rotator.  
Although O-star X-rays are generally
interpreted as emission from shock regions distributed throughout
the wind (e.g., Owocki, Castor, \& Rybicki 1988), X-ray emission
from $\theta^1$~Ori~C is not typical of most single, main-sequence
O~stars.  First, $\theta^1$~Ori~C is the only known O star
with large-amplitude, periodic X-ray variability.  
Second, $\theta^1$~Ori~C's peak X-ray activity level,
$L_{\rm X}/L_{\rm bol} \sim 1.8 \times 10^{-6}$, is higher 
(by a factor of five) than any other single O star detected in 
the {\it ROSAT} all-sky survey (Bergh\"ofer, Schmitt, \& Cassinelli 1996).
Third, the {\it ASCA} SIS spectrum of the Trapezium (Yamauchi et al. 1996) 
indicates very high temperature plasma with $T > 20$~MK.
Although the SIS cannot spatially resolve $\theta^1$~Ori~C 
from surrounding lower-mass stars, the HRI images and the PSPC
low-resolution spectra suggest that some of the high-temperature 
emission must be associated with $\theta^1$~Ori~C.
Conventional O-star shock models
do not predict sufficiently fast shocks to produce such hot plasma.
On the other hand, magnetically confined plasma (e.g, in coronal
loops on the Sun and other late-type stars) can be heated to
very high temperatures.

Assuming that X-rays from the magnetosphere are being absorbed
in the wind, X-ray variability would arise from varying wind absorption.
The amount of excess absorbing material
can be inferred from the \ion{C}{4} excess equivalent width
at high velocities around phase 0.5,
$W_{1548,1551} \sim 2.2$~\AA\ (Walborn \& Nichols 1994).
Assuming that the absorbing material is optically thin in \ion{C}{4},
we find a lower limit to the excess 
column density $N_{\rm C IV} \sim 3.7 \times 10^{14}$~cm$^{-2}$.
Howarth \& Prinja (1989) 
determine $N_{\rm C IV} \sim 9.4 \times 10^{14}$~cm$^{-2}$
from the P Cygni profile at low velocities.  If the wind and the
absorbing region have approximately the same abundance and ionization,
then 25--30\% of the column density measured at phase 0.5 
comes from the overlying absorption region. Since the high-velocity 
\ion{C}{4} absorption is not observed at phase 0.0, can this column 
density difference account for the observed X-ray variability amplitude?

In order to estimate the X-ray spectral shape of $\theta^1$~Ori~C,
we have analyzed the {\it ROSAT} PSPC spectrum of the Trapezium 
obtained in 1991 March over 4~d spanning phases 
0.36--0.63, i.e., near $\theta^1$~Ori~C X-ray minimum.
We have simulated {\it ROSAT} HRI count rates at phases 0.0 and 0.5
in XSPEC (Arnaud 1996) by fitting the PSPC spectrum obtained near phase 0.5
and varying the column density in the overlying absorption region.
The fit parameters are not well constrained, but the simulations 
suggest that a $\sim 25\%$ decrease in the
wind column density would result in a $\sim 40\%$ increase 
in the HRI count rate from phase 0.5 to 0.0. More detailed 
modeling of spatially resolved X-ray spectra is required, but our
preliminary estimates suggest that absorption in the
overlying, corotating wind of an oblique magnetic rotator
may be a viable mechanism for the observed X-ray variability.

\subsection{Magnetosphere Eclipses}

On an O-type magnetic rotator, X-ray variability might result from 
eclipses of the magnetosphere and/or from varying absorption in the 
overlying, corotating wind.
If the magnetosphere on $\theta^1$~Ori~C
extends out to many stellar radii like it does on $\sigma$~Ori~E
and if $i \sim 45^{\circ}$, then the X-ray variations
are not likely to arise from eclipses. 
However, if the magnetosphere
were closer to the stellar surface (i.e., 1--2~$R_{\star}$),
then eclipses could produce smooth X-ray variations.
This scenario can be tested with X-ray spectra obtained at
opposite phases.  If the X-ray spectra do not indicate
any appreciable change in the absorbing column density from X-ray 
minimum to X-ray maximum, then the variability is 
most likely due to eclipses.

\section{Challenges for the Oblique Magnetic Rotator Model}

The first two scenarios,
which require the presence of an unseen binary companion,
probably cannot account for the smooth, large-amplitude X-ray variations
seen on $\theta^1$~Ori~C.
Furthermore, there is little evidence that $\theta^1$~Ori~C
is a binary: the photospheric lines only show 
small, irregular radial-velocity variations (Stahl et al. 1996).  Such variability is 
characteristic of atmospheres of luminous O stars and is not strong evidence
for a companion (Bieging, Abbott, \& Churchwell 1989).
The period of the X-ray and H$\alpha$ emission is not
compatible with non-radial pulsations causing density fluctuations
in the O star wind.

The phase, period, and variability amplitude of the X-ray emission
on $\theta^1$~Ori~C do appear to be consistent with either of the last 
two scenarios. Like Stahl et al. (1996), we, too, conclude that the 
15-d period is the rotation period of the O star and that most of 
the observed variable emission and absorption phenomena can be 
explained if $\theta^1$~Ori~C is an oblique magnetic rotator.
Nonetheless, a few outstanding questions remain.

First, and most importantly, it would be useful to measure the
longitudinal magnetic field strength as a function of phase.
Establishing the existence and location of one or more magnetic
poles on $\theta^1$~Ori~C would represent the first definitive 
detection of magnetic fields on any O star.  
A surface magnetic field of a few hundred Gauss may be sufficient to produce 
wind corotation out to 10 stellar radii.  However, such a small field 
may be difficult to detect
because it would produce Zeeman shifts of only a few m\AA\ compared to
a polarized line width of $\sim 1$~\AA\ (Otmar Stahl, private communication).
Alternatively, the Hanle effect can be used to detect polarization 
differences in the
\ion{C}{4} $\lambda\lambda 1548, 1551$ resonance lines
for massive stars whose longitudinal $B$-field is less than 1~kG.
The Hanle effect refers to a change in the polarization of
resonantly scattered photons as a result of Larmor precession of the scattering
electron in a magnetic field
(see Cassinelli \& Ignace 1996).

Second, $v \sin i \sim 50$~km~s$^{-1}$, as measured from photospheric 
\ion{O}{3} $\lambda\lambda 3756, 3760$ line profiles, is too high.
If the 15.4~d period is the rotation period of the star,
then $v \sin i$ must be less than $v_{\rm eq} \sim 30$~km~s$^{-1}$,
assuming $R_{\star} \sim 8 R_{\odot}$ (Howarth \& Prinja 1989).
Non-rotational broadening mechanisms, which have not been taken into account 
in the $v \sin i$ determination of Stahl et al. (1996), may explain some
of the discrepancy. 
Possible broadening mechanisms include: stronger stark 
broadening due to higher gravity or atypically large
macroturbulence, related to the peculiar optical line-profile variations.
We note, however, that most main-sequence 
O stars show little evidence of macroturbulence (Ebbets 1979) and classical
Doppler broadening measurements are generally adequate for determining $v \sin i$ 
in slowly rotating early-type stars like $\theta^1$~Ori~C (Collins \& Truax 1995).

Finally, modeling of magnetospheric X-ray emission with non-spherically
symmetric wind geometries is essential.  Upcoming observations of the
Trapezium with {\it ASCA} at opposite phases of $\theta^1$~Ori~C may
help distinguish between the two competing hypotheses for the X-ray
variability observed by {\it ROSAT}.  Because of source confusion in
the Trapezium, though, a conclusive test may only be possible with
the {\it AXAF} CCD Imaging Spectrometer.

\acknowledgements

MG would like to thank Steve Skinner, Jeff Valenti, Peter Conti, and
Otmar Stahl for useful discussions about X-rays, magnetic fields, and O stars.
The authors wish to thank an anonymous referee for helpful suggestions.
This work was supported in part by NASA grants NAG5-2075 to the University of Colorado,
NAG5-1610 to the University of Georgia, and NAGW-2698 to the 
Harvard-Smithsonian Center for Astrophysics.

\newpage
\figcaption[figure_1.eps]{
{\it ROSAT} HRI lightcurve of $\theta^1$~Ori~C.
HRI count rates and $1\sigma$ uncertainties are plotted versus MJD 
for observations obtained 1995 Sep 1--21 (filled circles);
crosses indicate orbits with uncertain count rates (see \S 2).
The upper axis indicates the corresponding phase of $\theta^1$~Ori~C based 
on the ephemeris of Stahl et al. (1996).
The count rates and phases from 1991 Oct 2, 1992 Mar 22, and 1992 Sep 14
are plotted as open circles.
\label{figure_1}
}

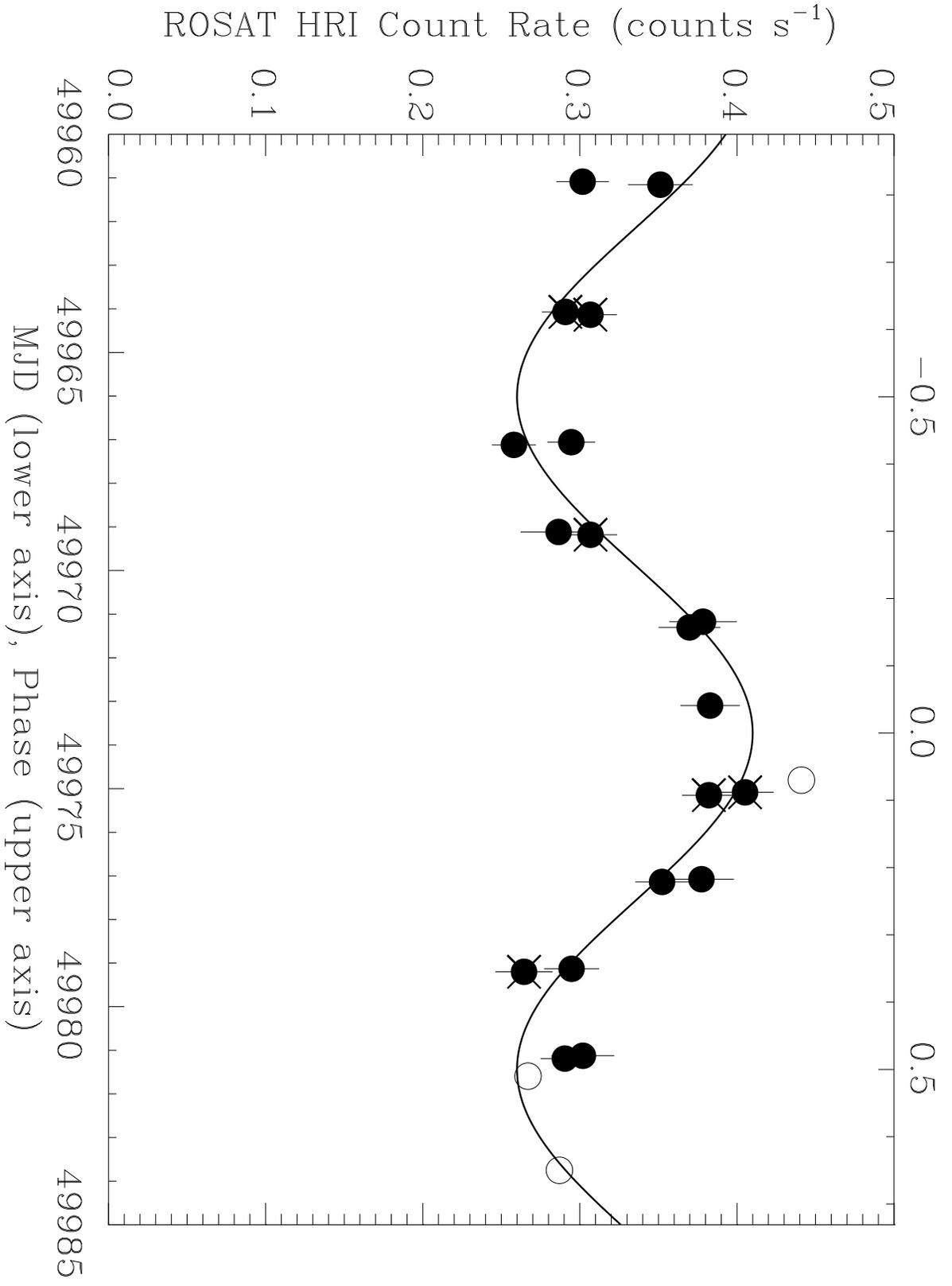
\end{document}